\documentstyle[epsfig,12pt]{article}
\def\singlespace {\smallskipamount=3.75pt plus1pt minus1pt
                  \medskipamount=7.5pt plus2pt minus2pt
                  \bigskipamount=15pt plus4pt minus4pt
                  \normalbaselineskip=15pt plus0pt minus0pt
                  \normallineskip=1pt
                  \normallineskiplimit=0pt
                  \jot=3.75pt
                  {\def\smallskip {\vskip\smallskipamount}}
                  {\def\medskip   {\vskip\medskipamount}}
                  {\def\bigskip   {\vskip\bigskipamount}}
                  {\setbox\strutbox=\hbox{\vrule
                    height10.5pt depth4.5pt width 0pt}}
                  \parskip 7.5pt
                  \normalbaselines}
\def\middlespace {\smallskipamount=5.625pt plus1.5pt minus1.5pt
                  \medskipamount=11.25pt plus3pt minus3pt
                  \bigskipamount=22.5pt plus6pt minus6pt
                  \normalbaselineskip=22.5pt plus0pt minus0pt
                  \normallineskip=1pt
                  \normallineskiplimit=0pt
                  \jot=5.625pt
                  {\def\smallskip {\vskip\smallskipamount}}
                  {\def\medskip   {\vskip\medskipamount}}
                  {\def\bigskip   {\vskip\bigskipamount}}
                  {\setbox\strutbox=\hbox{\vrule
                    height15.75pt depth6.75pt width 0pt}}
                  \parskip 11.25pt
                  \normalbaselines}
\def\doublespace {\smallskipamount=7.5pt plus2pt minus2pt
                  \medskipamount=15pt plus4pt minus4pt
                  \bigskipamount=30pt plus8pt minus8pt
                  \normalbaselineskip=30pt plus0pt minus0pt
                  \normallineskip=2pt
                  \normallineskiplimit=0pt
                  \jot=7.5pt
                  {\def\smallskip {\vskip\smallskipamount}}
                  {\def\medskip   {\vskip\medskipamount}}
                  {\def\bigskip   {\vskip\bigskipamount}}
                  {\setbox\strutbox=\hbox{\vrule
                    height21.0pt depth9.0pt width 0pt}}
                  \parskip 15.0pt
                  \normalbaselines}
\def\p{\partial}

\def\l{\left}
\def\r{\right}
\def\un{\underline}

\def\dis{\displaystyle}
\begin{document}
\thispagestyle{empty}
\begin{center}
{\huge Renormalisation Group Improved\\[10pt]  Thermal Coupling Constant\\[10pt] 
 In An External  Field}\\[40pt]
Avijit K. Ganguly  \\[25pt]
Centre for Theoretical Studies, IIsc Bangalore\\ Bangalore 560012 (India)\\[30pt]
\end{center}

\begin{center}
\small {Abstract}\\[10pt]
\end{center}

\begin{quote}
Starting from renormalised Effective Lagrangian, in the
presence of an external Chromo-Electric field at finite temperature, the
expression for thermal coupling constant ( $\alpha = \frac {g^2}{4 \pi}$ )
as a function of temperature and external field is derived, using
finite temperature two parameter renormalisation group equation of Matsumoto,
Nakano and Umezawa.
For some values
of the parameters, the  coupling constant is seen to  be approaching a value
 $\sim unity$.
\end{quote}

\vspace{10pt}

\noindent
Study of coupling constant in finite temperature field theory (FTFT)
{\cite{gross}} is of interest in various
disciplines of physics, for the study of  processes
taking place in different physical situations
 varying from Early Universe Cosmology, Astro-physics,
to Relativistic Heavy Ion Collision.
From the strong
interaction physics point of view, it has been extensively used
in the description of Quark Gluon Plasma, the elusive state of
matter, supposed to form in Relativistic Heavy Ion
collision.  Since the inception of the concept of asymptotic
freedom, in the context of non-abelian (NA) gauge theories,
it was argued in
ref{\cite{coll}},that, as in the high momentum
exchange processes, at high temperature/density or both of them,
strong coupling constant ($\alpha_R $) also decreases making
the system behave in an asymptotically free manner.  Since then,
from the point of
view of strong interaction physics, i.e. Quantum Chromo Dynamics
(QCD), many studies of the running coupling constant, have been
performed
 with different renormalisation schemes.\\

\noindent
In this note of ours, we, instead of taking full QCD into account,
and going into the details of those aspects,
will
be contented by showing the running of the strong coupling constant
with respect to temperature in the presence of an external
field with  a SU(2) or U(1) internal
symmetry present in it.
This model was originally studied by
 Schwinger{\cite{sch}} and 
has been applied successfully in RHIC {\cite{Avi1}} also in 
some astrophysical{\cite{adl}} as well as cosmological{\cite{jane}}
scenarios. \\

\noindent
In this note, we show that, in the presence of heat bath and an external
electric field, existence of a phase may be possible where 
the value of $\alpha_R$  can be around unity.
Infact depending
on the strength of the electric field, in the low temperature
region, the running coupling constant shows
an oscillatory behavior, where $\alpha_R(E,T)$ can have values
of the order of unity. But interestingly as the temperature  still increased
there exists
a critical "temperature"  
around which the coupling constant grows enormously and beyond
that it ( coupling constant ) changes its sign; perhaps signaling a
change of the phase of the system. Considering the limitations of this
study, apriori it is not clear whether this analysis, 
can be extended to that domain of temperature, however we will discuss this
point as we go along.\\

\indent
Having stated the motivation behind this study, we go about 
explaining the organisation of this document. 
That  is as follows, in section II
we provide the necessary  background for this study, in section III
we give the essential steps to get to the expression of the renormalised
effective action. In section IV the computation of the coupling
constant with the numerical results will be presented.
In the last section we conclude by describing the future plans in
this direction.

{\section {\bf Thermal running coupling constant from Effective action}} 

\indent
A formalism for the computation of the running coupling constant, at zero
temperature, was developed by Coleman and Weinberg (CW){\cite{cole}}
in the context of studying spontaneous symmetry breaking,
in gauge theory, by radiative correction.
For the purpose of computing the same quantity at nonzero temperature,
one needs to 
generalise the method of CW to finite temperature by using
the proper renormalisation group equations.
At finite temperature, following ,  Matsumoto,
Nakano and Umezawa{\cite{umezawa}} we work with the two
parameter renormalisation (RG) equations at finite temperature.
In this formalism, in addition to the momentum variable $\mu$ one introduces a
dimensionful parameter $\tau$ at the corresponding renormalisation
point (at nonzero temperature) and then writes a set of renormalisation group
equations (RGE)
w.r.t $\tau$. From the solution of these set of( two parameter )
RGEs one recovers the corresponding flow of the coupling
constant w.r.t temperature or external field or both.\\

\noindent
Since  the effective action is the generating functional  of the 
one particle irreducible (1pI)
Green functions, the RG equation satisfied by the Green functions
should also be satisfied by the ${\cal L}_{eff}$. So following the
authors of ref {\cite{umezawa}} the      
 flow equations can be written in terms of the 
effective Lagrangian as 
\begin{equation} 
\l( {\mu {\frac {\partial}{\partial \mu}} + \beta_{\mu} {\frac {\partial}{\partial g}} +2
\gamma_{\mu}(g)} {\cal F} {\frac {\partial} {\partial \cal F} }\r) { {\cal L}_{eff}}  = 0
\end{equation} 
\begin{equation}
\l( {\tau {\frac {\partial}{\partial \tau}} + \beta_{\tau} {\frac {\partial}{\partial g}} +2
\gamma_{\tau}(g)} {\cal F} {\frac {\partial} {\partial \cal F} }\r) { {\cal L}_{eff}}  = 0
\end{equation}
Here the effective Lagrangian i.e. ${\cal L}_{eff}$={\bf ${\cal
L}_{eff}(gE,T,\mu, \tau)$ } and ${\cal F}$ corresponds to
${(E^2-B^2) \over{2}}$; though for our purpose we set $B=0$ 
We rewrite these equations (1) and (2) in terms of
dimensionless quantity $ \bar L_{eff} = \frac {\partial {\cal L}_{eff}}{\partial
\cal F} $ so as to get
\begin{equation} 
\l( {\mu {\frac {\partial}{\partial \mu}} +{ \beta_{\mu}} {\frac {\partial}{\partial g}} +2
{ \gamma_{\mu}(g)}}\l(1 + {\cal F} {\frac {\partial} {\partial \cal F} }\r)\r) { {\bar L}_{eff}}  = 0
\end{equation} 
\begin{equation}
\l( {\tau {\frac {\partial}{\partial \tau}} +{ \beta_{\tau}} {\frac {\partial}{\partial g}} +2
{ \gamma_{\tau}(g)}} \l(1 + {\cal F} {\frac {\partial} {\partial \cal F} }\r) \r) { {\bar L}_{eff}}  = 0
\end{equation}
\noindent
Here,$\beta_{\mu}$
$\gamma_{\mu}$
$\beta_{\tau}$,
are $\gamma_{\tau} $ are defined as
$\beta_{\mu}= \mu {\frac {d g}{d \mu}} $
$\gamma_{\mu}= \mu {\frac {d ln Z}{d \mu}} $ and 
$\beta_{\tau}= \tau {\frac {d g}{d \tau}}  $,
$\gamma_{\tau}= \tau{\frac {d ln Z}{d \tau}}. $

\indent 
Subtracting equation (3) from equation (4) and introducing
another new variable $ \zeta $ ,such that $ ln {\frac {\mu}{\tau}} =
ln {\zeta} $,we rewrite the corresponding equations to arrive at
\begin{equation}
\l( {\zeta {\frac {\partial}{\partial \zeta}} +{ \beta_{\zeta}} {\frac {\partial}{\partial g}} +2
{ \gamma_{\zeta}(g)}} \l( 1+ {\cal F} {\frac {\partial} {\partial \cal F} }\r) \r)
{ {\bar L}_{eff}}  = 0
\end{equation}
At this stage, 
if we introduce a dimension less quantity $ \kappa = 2 ln \l[ \frac
{2 {\cal F}^{1/2}} {\zeta} \r] $,  equation (5) takes the form
\begin{equation}
\l(-{ {\frac {\partial}{\partial \kappa}} +{\bar \beta_{\kappa}} {\frac {\partial}{\partial g}} +2
{\bar \gamma_{\kappa}(g)}} \r)
{ {\bar L}_{eff}}  = 0
\end{equation}
\indent
The quantities $\bar \beta_{\kappa}$ $\bar \gamma_{\kappa}$ are defined to be
$\bar \beta_{\kappa} = {\beta_{\kappa} \over {(1-\gamma_{\kappa}})}$ and
$ \bar \gamma_{\kappa} = {\gamma_{\kappa} \over {(1-\gamma_{\kappa}})}$
This is the basic equation, that when solved  with the boundary condition,
\begin{equation}
 \l( {\frac {\partial {\cal L}_{eff}}{\partial \cal F}} \r)_{ \kappa =0} = 1
\end{equation}
will provide us with the information, how the coupling constant changes
with the change of the corresponding scale defined by the
strength of the  external field and/or temperature.  
This boundary condition is chosen at a particular scale, such that;
$ \l[ {\frac {2 gE T}{\mu}} \r] =1 $. The method of  solving equation (6) quite standard,
and the solution is,

$$   {{\bar L_{eff}} \l(\kappa, g \r)} = - exp { \l[ 2 {\int_0}^{\kappa} 
\bar \gamma \l( \bar g(x,g ) \r) d x  \r]} $$
as can be checked by direct substitution also. 
Since $ \beta = - g \gamma $ 
( This can be derived from the relation between the bare and the renormalised
coupling constant i.e. 
$ g_b = Z_r g $ and differentiate it w.r.t $ \kappa $, then set the right side
to zero and use standard results )
and as  $$ \frac {d \bar g ( \kappa )}{d \kappa} = \beta \l( g \l(
\kappa \r) \r) $$ \\
using these two results one can arrive at the relation

\begin{equation}
{\bar {\ L}_{eff}} = \frac {- 1}{ \bar g^2( \kappa ) }
\end{equation}
\noindent
Using the relation, $  \bar L_{eff} = \frac {\partial {\cal L}_{eff}}{\partial \cal
F} $, as defined before, the equation for the flow of running coupling constant
becomes
 
\begin{equation}
 \bar L_{eff} = \frac {\partial {\cal L}_{eff}}{\partial \cal F}
= \frac {- 1}  { \bar g^2( \kappa ) }
\end{equation}
We will employ this relation to see the flow of coupling at
finite temperature and nonzero external field.
  
\noindent
{\section {\bf Renormalisation of the Effective Lagrangian}}

We start from the ``partition function" in Minkowski space defined
by
\begin{equation}
Z[A] = \frac{\int D {\bar{\psi} }D \psi e^{i \int L d^4 x}}{\int D \bar{\psi} D
\psi e^{i \int L_o d^4 x}}
\end{equation}
where $L = \bar{\psi} \l( i  \gamma_{\mu} {\partial}^{\mu} - g
\gamma_{\mu}{A_a}^{\mu} \tau_a \r) \psi - m \bar{\psi} \psi$ is the
fermionic Lagrangian in the presence of external vector field
${A_\mu}^a$ with a $SU(2)$ color symmetry present in it, $\tau_a$'s are the Pauli matrices
and  $L_o =
\bar{\psi} \l( i \gamma_{\mu} {\partial}^{\mu} - m \r) \psi$ is the free fermionic
Lagrangian
so that $Z[0] = 1$.  

\indent

Since we are interested in evaluating the effective action, 
in the presence of external
chromo-electric field, we choose $ {A_0}^a = -E^a z $
and other components of A to be equal to zero. 
The expression for the effective action is calculated, from the euclidean
action,
\begin{equation}
 S_\beta = \frac{1}{\beta^2} 
\sum^\infty_{n = - \infty}
\int \bar{\psi}_n(x) \l[ \l( \omega_n \gamma^0 + g {A^a}_{o} \tau^a \gamma^0
\r) + i \gamma^j \partial_j - m \r] \psi_n(x) d^3 x 
\end{equation}
that is gotten after wick rotating the time component of
 the original Minkowski space action and 
compactifying the imaginary time direction over a length scale
 $\beta= {\frac{1}{T}}$ ( for details  see
ref \cite {gross}). 
Here the fermion fields, i.e the $ \psi $ 's are in the
fundamental representation of SU(2) defined as  
$ \psi (x) = {\frac {1}{\beta}} {\sum_n} \int \frac {d^3 x}{{2\pi}^3} e
^{-i \l(\omega_n \tau - p x\r) } {\tilde\psi_n} \l(\omega_n, p \r)
$  and
$$
S_{o \beta} = \frac{1}{\beta^2} 
\sum^\infty_{n = - \infty}
\int \bar{\psi}_n(x) [ \omega_n \gamma^0 + i \gamma^j {\p}_j - m ]
\psi_n(x) d^3 x 
$$
Then one arrives at  the finite temperature
partition function  from the Euclidean action $ S_{\beta}
$ as
\begin{equation}
Z \l[ A \r] = \frac{\prod_{n=- \infty}^{ \infty}\int D \bar{\psi}_n D {\psi}_n
e^{-S_\beta}}{\prod^{\infty}_{n= - \infty} \int D \bar{\psi}_n D \psi_n 
e^{-S_{o \beta}}}
\end{equation}

\noindent
From this partition function one arrives at the expression for the
effective Lagrangian,( details will be reported elsewhere ref{\cite{Avi2}}. )
\begin{equation}
\begin{array}{lllll}
{\cal{R}}e{\cal{L}^Q}_{eff} &=& - \frac{1}{4 \pi^2} p.v \int^\infty_o
\frac{ds}{s^3} e^{-sm^2} \l[ (gEs) cot (gEs)-1 \r] - \frac{1}{2 \pi^2}
\Sigma^\infty_{n=1} (-1)^n \left[ p.v \int^\infty_o \frac{ds}{s^3} 
\right.\\ [8pt]
&&  \l[ \cosh \l(
n g \beta \bar{A}_o \r) (gEs) cot (gEs) cos \l.( n^2 \beta^2
gE/4 \r.)
- 1 \r]  e^{{-sm^2} - \frac{n^2
\beta^2}{4s}}  \\[8pt]  && \left. + \pi \Sigma^\infty_{l=1}\cosh \l(
n g \beta \bar{A}_o \r) \l({\frac
{gE}{l \pi}}\r)^2 sin \l( {\frac{n^2 \beta^2 gE}{4}} \r)
e^{-{\frac {n^2 \beta^2 gE}{4 \pi l}} - {\frac {m^2 \pi l}{gE}}}
 \r]
\end{array}
\end{equation}         
with $\bar{A}_o = - E z$.

\indent
As one can see that the divergence of this effective
Lagrangian is contained in the $ T=0 $ piece; following
Schwinger one can remove this divergence by subtracting out
a quantity ( that is found by analysing the $ s \rightarrow 0 $ behavior
of the zero temperature piece of the effective Lagrangian) defined
as
\begin{equation}
{\cal L}_{d} =- {\frac {(gE)^2}{12 \pi^2}} {\int_0}^{\infty} {\frac {ds}{s^3}
 e^{-sm^2}}
\end{equation}
In order to get a finite answer, from this divergent ${\cal L}_{eff}$
one absorbs these infinities in the redefination of the field and
the coupling constant, by redefining them  in terms of some
multiplicative renormalisation constant such that $ g_R E_R =
g_{un} E_{un} $ where subscript $ R/Un $ refers to 
whether the quantity is renormalised
or unrenormalised. 
\begin{equation} 
g_R={g_{un} \over { (1+Z)^{1/2}}} 
~~~and~~~~ 
E_R=  (1+Z)^{1/2} E_{un}
\end{equation}
Here Z is a function of $L_d$.
So with this redefination of the field and the coupling constant
the real part of the effective Lagrangian can be expressed as 

\begin{equation}
\begin{array}{lll}
{\cal{L}}_{eff} &=&{\frac {E^2}{2}}-\l[ \frac{1}{4 \pi^2} \int^\infty_o
\frac{ds}{s^3} e^{-sm^2} \l[ (gEs) cot (gEs)-1+{(gEs)^{2} \over 3} \r]
+ \frac{1}{2 \pi^2}
\Sigma^\infty_{n=1} (-1)^n \l.[\int^\infty_o \frac{ds}{s^3} 
\r. \right.\\ [8pt]
&& \l. \left. \times \l[ \cosh \l(
n g \beta \bar{A}_o \r) (gEs) cot (gEs) cos \l.(  n^2 \beta^2
gE/4 \r.)
- 1 \r]  e^{{-sm^2} - \frac{n^2
\beta^2}{4s}} \r. \r. \\  
&& \l. \l. +~ \pi \Sigma^\infty_{l=1} \cosh \l(
n g \beta \bar{A}_o \r) \l({\frac
{gE}{l \pi}}\r)^2 sin \l( {\frac{n^2 \beta^2 gE}{4}} \r)
e^{-{\frac {n^2 \beta^2 gE}{4 \pi l}} - {\frac {m^2 \pi l}{gE}}}
\r] \r]
\end{array}
\end{equation}         
The first term here is the tree level term, and since this satisfies
the R.G equation trivially, the quantity ${\cal{L}}_{teff}$ also
satisfies the R.G equation.
 As a next step we will scale $E$
 $: \rightarrow {\frac {E}{g}} $  to get
\begin{equation}
\begin{array}{lll}
{\cal{L}}_{eff} &=&{\frac {E^2}{2 g^2}}-\l[ \frac{1}{4 \pi^2} \int^\infty_o
\frac{ds}{s^3} e^{-sm^2} \l[ (Es) cot (Es)-1+{(Es)^{2} \over 3} \r]
+ \frac{1}{2 \pi^2}
\Sigma^\infty_{n=1} (-1)^n \l.[\int^\infty_o \frac{ds}{s^3} 
\r. \right.\\ [8pt]
&& \l. \left. \times \l[ cosh \l(
n  \beta \bar{A}_o \r) (Es) cot (Es) cos \l.( n^2 \beta^2 E/4 \r.)
- 1 \r]  e^{{-sm^2} - \frac{n^2
\beta^2}{4s}} \r. \r. \\  
&& \l. \l. +~ \pi \Sigma^\infty_{l=1}cosh \l(
n  \beta \bar{A}_o \r)  \l({\frac
{E}{l \pi}}\r)^2 sin \l( {\frac{n^2 \beta^2 E}{4}} \r)
e^{-{\frac {n^2 \beta^2 E}{4 \pi l}} - {\frac {m^2 \pi l}{E}}}
\r] \r]
\end{array}
\end{equation}         

\noindent
Now employing equation (9) we finally come to the expression for
running coupling constant at any temperature or electric field defined
as 
\begin{equation}
{1 \over {4 \pi^2 {g_R}^2}}={1 \over {4 \pi^2 {g_o}^2}} -\l[ DF_1 +
DF_2 + DF_3 \r]
\end{equation} 
where the quantities ${DF_1}$ and $DF_2$ are the finite temperature
pieces and $DF_3$ is the zero temperature piece.
These quantities are defined as 
\begin{eqnarray} 
DF_1 & = & \dis\frac {1} {\l(2\pi^2 E^2\r)} \dis\sum_{n=1}^{\infty} (-1)^n
\l[\l[ (n\beta  {\bar A}_0) sinh (n\beta  {\bar A}_0) 
cos \l(\dis\frac {n^2\beta^2 E} {4}\r ) \r. \r. \nonumber \\ 
&& \l.\l. - \l(\dis\frac {n^2\beta^2
E} {4}\r ) cosh (n\beta  {\bar A}_0) 
sin \l(\dis\frac {n^2\beta^2 E} {4}\r ) \r.\r. \nonumber \\
&& \l.\l. + ~~ cosh (n\beta  {\bar A}_0) 
cos \l(\dis\frac {n^2\beta^2 E} {4}\r ) \r] \dis\int^\infty_0
\dis\frac {ds} {s^3} (Es~cot(Es)) 
e^{-s m^2 - \frac {n^2\beta^2} {4s}} \r. \nonumber \\
&& \l. - ~ cosh (n\beta  {\bar A}_0) 
cos (n^2\beta^2 E/4)  \dis\int^\infty_0
\dis\frac {ds} {s^3} \dis\frac {(Es)^2} {sin^2(Es)}
e^{-s m^2 - \frac {n^2\beta^2} {4s}} \r] \\
%
& & \nonumber \\
& & \nonumber \\
& & \nonumber \\
DF_2 & = & \dis\frac {1} {\pi } \dis\sum_{n=1}^{\infty} (-1)^n
\dis\sum_{\ell = 1}^{\infty} \l[ \l( \dis\frac {1} {\ell \pi}
\r)^2 e^{-\frac {m^2\pi\ell} {E} - \frac {n^2\beta^2 E}
{4\pi\ell}} \r] \l[ cosh(n\beta  {\bar A}_0) sin \l(\dis\frac {n^2\beta^2
E} {4}\r ) \r. \nonumber \\
&& + \l. \l( \dis\frac {n\beta  {\bar A}_0} {2} \r) sinh(n\beta
 {\bar A}_0) sin \l(\dis\frac {n^2\beta^2 E} {4} \r) \r.
\nonumber \\
&& + \l. \l(\dis\frac {n^2\beta^2
E} {8}\r) cosh(n\beta  {\bar A}_0) cos\l(\dis\frac {n^2\beta^2
E} {4} \r) \r. \nonumber \\
\vspace{4mm}
&& + \l. \dis\frac {1} {2} cosh(n\beta  {\bar A}_0) sin
\l(\dis\frac {n^2\beta^2 E} {4} \r) 
\l(\dis\frac {\pi\ell m^2} {E}  - \dis\frac {n^2\beta^2
E} {4\pi\ell} \r) \r] \\
& & \nonumber \\
& & \nonumber \\
& & \nonumber \\
DF_3 &=&  {\frac{1}{4 \pi^2 E^2 }} \int^\infty_o
{\frac {ds}{s^3}} e^{-sm^2} \l[ (Es) cot(Es)-{(Es)^2 \over sin^2(Es)}+{2(Es)^{2}
\over 3} \r]
\end{eqnarray}

\noindent
Using these expressions the running coupling constant is defined
as ( at some temperature T and electric field E)
\begin{equation}
\alpha_R \l( E, T \r) = \dis\frac {\alpha_o}{1 - 4 \pi^2 {g_o}^2 \l[ DF_1
+DF_2+DF_3 \r]}
\end{equation}
Here $\alpha_0$ is the strong coupling constant at the renormalisation
point.

{\section {\bf Analysis Of The Result}}

\indent
Equation (22) provides one with the relation, how the
coupling constant varies with the external parameters present in
the theory. A close inspection, of the aforementioned relation
shows that if the denominator, on the right hand side of the equation
becomes less than $one $ and tends to zero, the
value of $\alpha_R$ increases. In view of the fact that the expression
$DF_1$ is has terms with a relative sign difference between them,
it might be justified to imagine that for some value of the temperature
or the electric field the denominator may become less than one.
Since the terms there are highly oscillating, it is difficult
to come  to a compact analytical expression for them. In
any case we have tried to evaluate them numerically 
 and  have plotted
the value of the corresponding coupling constant with the
 variation in the temperature;(the corresponding $\alpha_R (E,0) \sim 0.1 $) .
There are several features that emerges from this analysis
namely;
i)
the magnitude of the coupling constant seems to peak around $\sim$ unity
or even more, for some value of the  
temperature,
ii)for distances away from the center (i.e. $ A_o \neq 0 $), even though it
(thermal coupling constant) continues to 
peaks around the same value of the temperature but it's magnitude increases.   
 A qualitative
interpretation of this phenomena may be given like this; at zero
temperature, in the presence of an 
 external electric field the particles and the antiparticles will try to
  move away from 
each other {\cite {chodos}} and naturally the coupling constant appears
 to be smaller in magnitude.
 But at finite temperature, because of the
 thermal pressure( this balances the force due to the external
field), 
the particles and the antiparticles are held close to each
other in a metastable state and hence  the magnitude of $\alpha_R(E,T)$ appears
to have increased.\\

\indent
The space dependence of $\alpha_R(E,T)$ is consistent with the fact that,
 at finite temperature, the Effective Lagrangian it self is position dependent.
It is worth remembering that, the particle  production rate -
as follows from the imaginary part of the Effective Lagrangian -  increases
as one moves away from the center of the system, therefore the number density
of particles, away from the center, are more. Other than thermal pressure,
this effect additionally contributes to hold the
partice, anti-partcle  pairs, close to each other,
hence there is an increase in $\alpha_R(E,T)$ at distances d $\ne$ 0 
than the same at $d = 0$.  
Any further increment of the temperature, shows the existence
of a critical temperature `$T_c$' ( it depends on the magnitude of other parameters
present in the theory ), around which $\alpha_R(E,T)$  grows
enormously and beyond that it changes its sign. 
We feel  that around this temperature, one is ought to take
the quantum back reaction in to account.
This effect has been studied before\cite {brout} at zero temperature,
where the external field was seen to admit a complex value.

{\section {\bf Conclusion }}        
 
\indent

 We believe
that there might be some
observational consequence of this phenomena in the
physical situations considered earlier, for instance
during the production part
of QGP in RHIC as well as studying properties of mesons 
in a hot medium \cite{pasu}. It is also reasonable to assume that, this can be
tested in the future Heavy Ion Experiments to be held at
CERN and BNL. Lastly it will also be  of interest to compute
the screening of the external field in the plasma, taking
the thermal coupling constant into account. Some work along
this direction is under progress, and will be reported elsewhere.\\    
   
\noindent
{\bf Acknowledgements}

The author would like to thank A.D. Patel for going through the manuscript
and making useful suggestions.

\newpage

\noindent {\Large \bf Figure Captions: }

\vspace{0.8cm}

\noindent {\bf Fig. 1 }: Behavior Of Running Coupling
Constant with Temperature. Distance= 0.1 ~fm, Mass= 1.0 GeV,
, gE=1.15 $~GeV^{2}$ and $\alpha_0(0,0)~=~0.3$\\

\noindent {\bf  Fig. 2} : Behavior Of Running Coupling
Constant with Temperature. Distance= 0.0 ~fm, Mass= ~1.0 GeV,
, gE=1.15~$GeV{^2}$ and $\alpha_0(0,0)~=~0.3$.
\newpage
\begin{figure}
\begin{center}{\epsfig{file=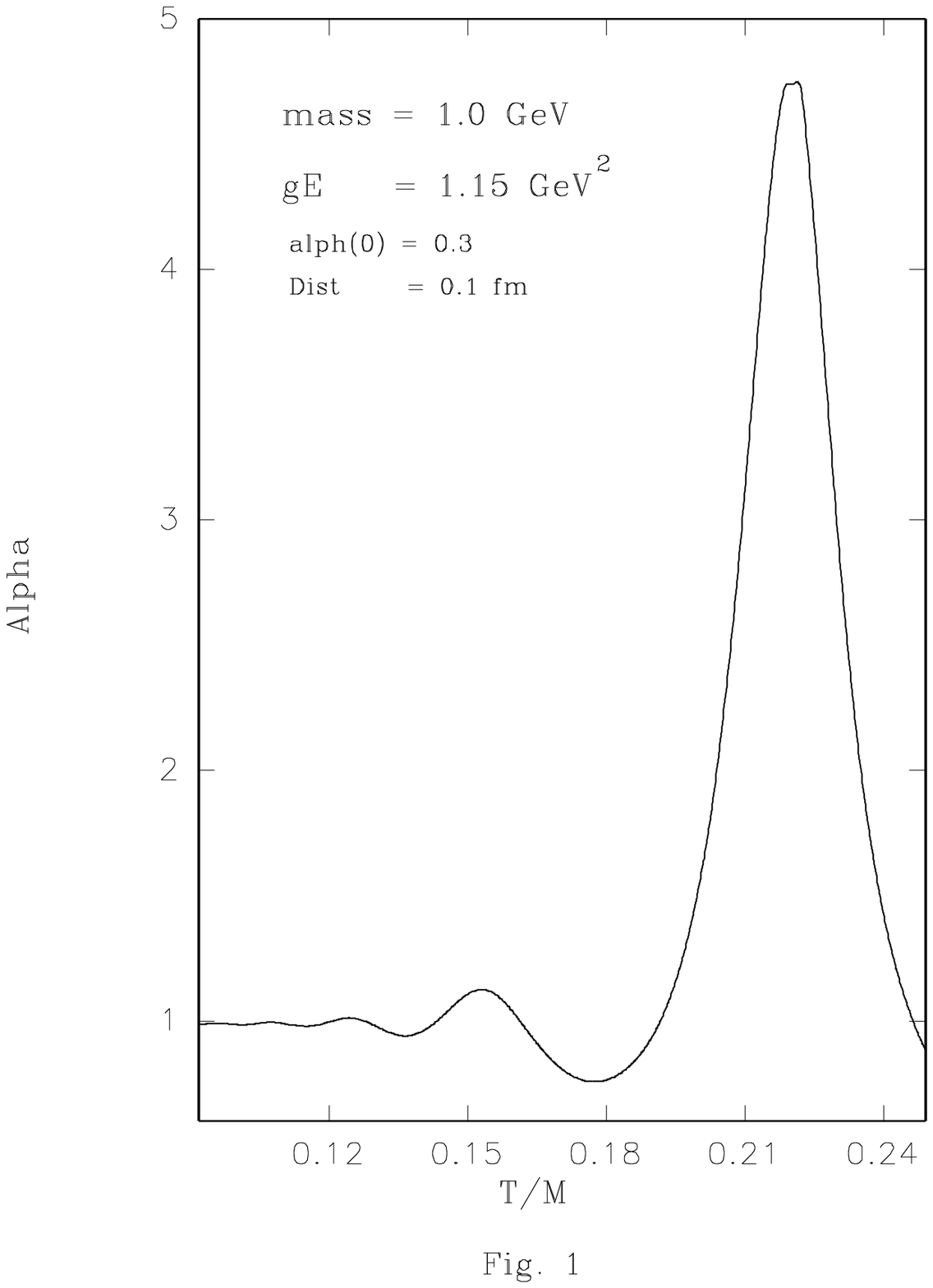, width=400pt}}\end{center}
\end{figure}
\newpage
\begin{figure}
\begin{center}{\epsfig{file=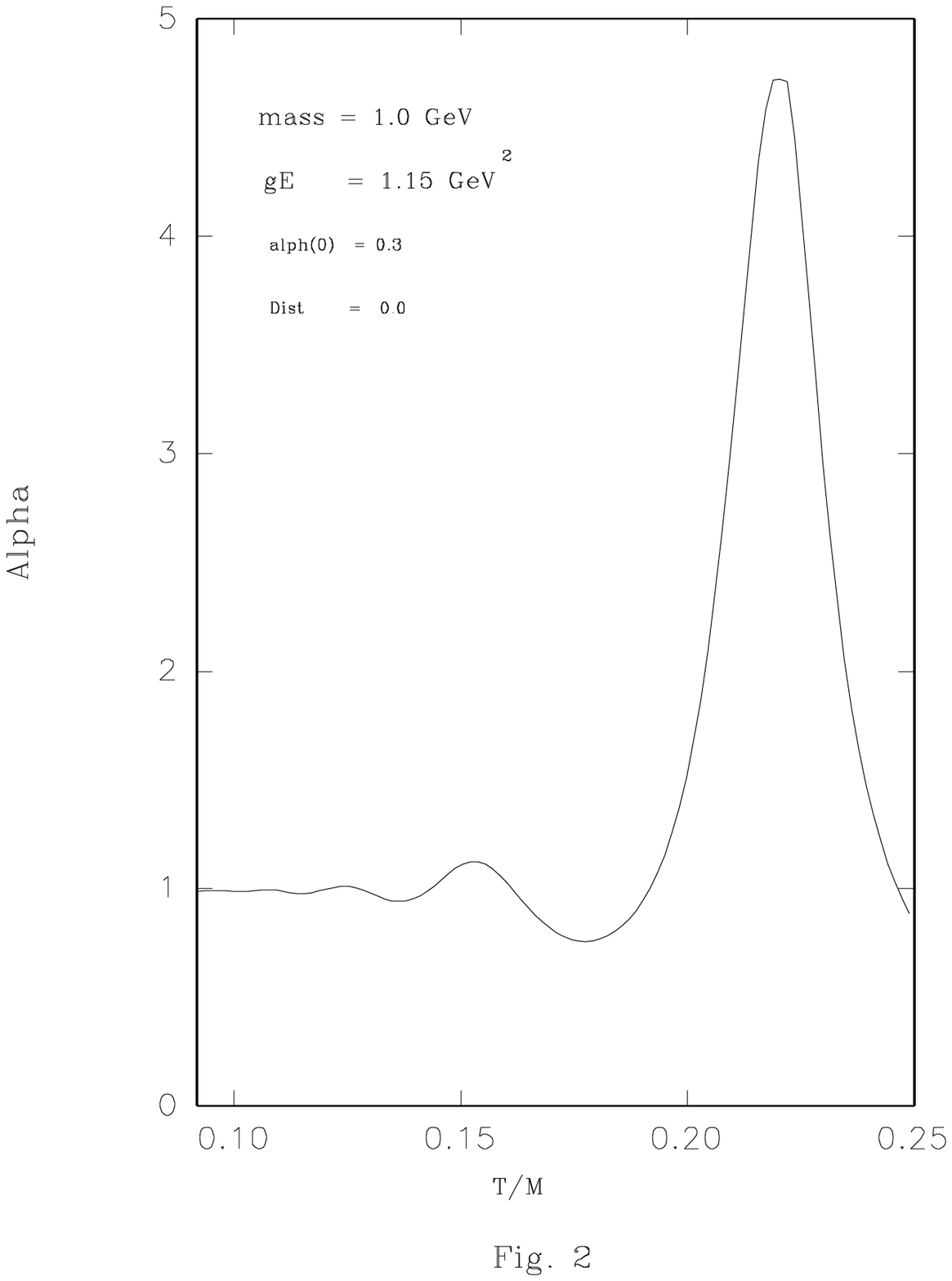,width=400pt}}\end{center}
\end{figure}

\noindent
\begin{thebibliography}{9}
\bibitem{gross} D.J.Gross, R.D. Pisarski, and L. G. Yaffe, Rev.
Mod. Phys. \un{53},(1981),43;\\
A. Niemi and G. W. Semenoff, Nucl. Phys. \un{B230} (1984), 181;\\
N.P. Landsman and Ch. G. van Weert, Phys. Rep. 145,(1987),141;\\
L. Dolan and R. Jackiw, Phys. Rev. \un{D9}, (1979), 3312;\\
J. Kapusta, Finite-Temperature Field Theory (Cambridge
University Press, 1989).


\bibitem{coll} J.C. Collins and M. J. Perry, Phys. Rev.Lett. \un{34},(1975),
1353.

\bibitem{sch} J. Schwinger Phys. Rev.\un{D82}; (1951),664.


\bibitem{Avi1} A.K. Ganguly, P.K. Kaw and J.C. Parikh, 
Phys. Rev. \underline{C51} (1995) 2091;\\
A.K. Ganguly, P.K. Kaw and J.C. Parikh,  Phys. Rev. \un{D46},
(1993), R2983.\\
 F.E. Low, Phys. Rev. \un{D12}, (1975),163\\
S. Nussinov, Phys. Rev. Lett. \un{34},(1975), 1286.\\
C.S. Warke and R.S. Bhalerao, Pramana J. Phys. \un{38}, 37
(1992) and references therein.

\bibitem{adl} J.K. Daugherty and I. Lerche, Phys. Rev. \un{14},(1976),340,\\
T. Damour and R. Ruffini, Phys. Rev. Lett. \un{35}, (1975), 1975.\\
S. L. Adler, Ann. Of Phys. \un{67}, (1971),599\\
P. Elmfors, D. Persson and B.S. Skagerstam, Phys. Rev. Lett. \un{71},(1993),480 .
 
\bibitem{jane} Pijushpani Bhattacharjee, Jan-e Alam, Bikash
Sinha, Sibaji Raha, Phys. Rev. \un{D48}, (1993), 4630 and references therein.


\bibitem{chodos} Alan Chodos, Andras Keiser and David A. Owen,
Phys. Rev. \un{D50}, (1994), 3566;\\
Alan Chodos, David A. Owen and C. M. Sommerfield, Phys. Lett. \un{B212},
(1988),491. and references therein \\ 
Alan Chodos, K. Everding and David A. Owen Phys. Rev. \un{D42},
(1990), 2881.

\bibitem{Avi2} Avijit K. Ganguly:  To be submitted.   

\bibitem{cole} S. Coleman and E. J. Weinberg, Phys. Rev. \un{D7},(1973),1888.

\bibitem{umezawa} H. Matsumoto, Y. Nakano and H. Umezawa, Phys. Rev.\un{D29},
 (1984), 1116.

\bibitem{brout} R. Brout, S. Massar, S. Popescu, R. Parentani, Ph. Spindel
  
Quantum Back Reaction On a Classical Field, ULB-TH 93/16,UMH-MG 93/03(and
references therein).  

\bibitem{pasu}  J.~Pasupathy,Mod. Phys. Lett.A12,(1997),1943; nucl-th/9703015.

\end{thebibliography}
\end{document}